\documentclass[%
 amsmath,amssymb,
 aps,
prd,
]{revtex4-1}

\usepackage{graphicx}
\usepackage{dcolumn}
\usepackage{bm}

\begin{document}

\title{Classification of black holes in three dimensional spacetime by the $W_{1+\infty}$ symmetry}
\author{Jingbo Wang}
\email{ shuijing@mail.bnu.edu.cn}
\affiliation{Institute for Gravitation and Astrophysics, College of Physics and Electronic Engineering, Xinyang Normal University, Xinyang, 464000, P. R. China}
 \date{\today}
\begin{abstract}
The BMS symmetry and related near horizon symmetry play important roles in holography in asymptotically flat spacetimes. They may also be crucial for solving the information paradox. But we still don't fully understand those infinite-dimensional symmetries. On the other hand, $W_{1+\infty}$ symmetry is the quantum version of area-preserving diffeomorphism of a plane. It is a dynamical symmetry of quantum Hall liquid and can be used to classify the quantum Hall universality classes. In this paper, we will show that the near horizon symmetry can be obtained from the $W_{1+\infty}$ symmetry. Based on this result, the black holes in three dimensional spacetime can be classified just as in the quantum Hall liquid. It also gives the result that the radius of black hole are quantized. This gives another evidence that our early claim ``black hole can be considered as a kind of topological insulator" is correct.
\end{abstract}
\maketitle
\section{Introduction}
The Bondi-Metzner-Sachs(BMS) group \cite{bms1,bms2,bms3} play important roles in holography in asymptotically flat spacetimes \cite{flat1,flat2,flat3,flat4,flat5}. The corresponding algebra is an infinite-dimensional extension of the translation part of the Poincare algebra. Near the black hole event horizon, there also exist a similar infinite-dimensional symmetry algebra \cite{nhs3,nhs4,nhs2,nhs1,nhs5}, which may be used to solve the information paradox.

Recently the author claimed that ``black hole can be considered as a kind of topological insulator" \cite{wangti1,wangti2}, and two evidences were given to support this claim. The first evidence comes from the black hole ``membrane paradigm" \cite{mp1,mp2}, which says that the horizon of black hole behaves like an electrical conductor and the vacuum inside can be considered as an insulator. The second evidence comes from the fact that the horizon of black hole can support massless modes \cite{whcft1,wangti2}. Those are two key properties of topological insulator. Since the infinite-dimensional symmetry algebra appears in the black hole side, it should also appear in the topological insulator side. And actually it does!

In $2+1$ dimensional spacetime, the topological insulator \cite{ti1,ti2} (also called quantum spin Hall state) can be realized as a bilayer quantum Hall system with opposite $T-$symmetry \cite{tibf1}. For quantum Hall effect, a fundamental property is that the electrons form an incompressible fluid \cite{incom1,incom2}, that is, the droplets of incompressible fluid have constant area. The small deformations of droplet at constant area can be generated by re-parametrizations of the coordinates of the plane with unit Jacobian, i.e. the area-preserving diffeomorphisms \cite{area1,ctz1,ctz2,walg1}. Those diffeomorphisms form an infinite-dimensional group, and the revelent algebra is called $w_\infty$ algebra. A quantum version of this algebra is called $W_{1+\infty}$ algebra. This $W_{1+\infty}$ algebra was used to classify the universality classes of quantum Hall incompressible fluids \cite{ctz2}.

The paper is organized as follows. In section II, it is shown that the near horizon symmetry algebra in three-dimensional spacetime is a subalgebra of  $W_{1+\infty}$ algebra. In section III, based on the $W_{1+\infty}$ algebra, black holes are classified just as in quantum Hall liquid. Section IV is the conclusion.
\section{Near horizon symmetry from $W_{1+\infty}$ symmetry}
In this section we consider the near horizon symmetry in three-dimensional spacetime from the $W_{1+\infty}$ symmetry. The algebra of near horizon symmetry is given in Ref.\cite{nhs1} and we just summarize the main results. The near horizon symmetry algebra is a semidirect sum of the Witt algebra generated by $Y_n$ with an abelian current $T_n$, and the commutation relations read
\begin{equation}\label{3}\begin{split}
  [T_m, T_n]&=0,\\
  [Y_m, T_n]&=-nT_{m+n},\\
  [Y_m, Y_n]&=(m-n)Y_{m+n}.
\end{split}\end{equation}
The $T_n$ generates a supertranslation and $Y_n$ generates a superrotation. Through the Sugawara construction one can get the $\mathfrak{bms}_3$ algebra.

In the following we will show that the near horizon symmetry algebra (\ref{3}) is a subalgebra of the $W_{1+\infty}$. The generators of the $W_{1+\infty}$ algebra $L^{(n)}_m$ satisfying the following relation \cite{walg1},
\begin{equation}\label{4}\begin{split}
  [L^{(0)}_m, L^{(0)}_n]&=0,\\
  [L^{(1)}_m, L^{(0)}_n]&=-n L^{(0)}_{m+n},\\
  [L^{(1)}_m, L^{(1)}_n]&=(m-n)L^{(1)}_{m+n},\\
  [L^{(2)}_m, L^{(0)}_n]&=-2n L^{(1)}_{m+n},\\
  \cdots.
\end{split}\end{equation}
It is easy to see that, after the identification $T_m=L^{(0)}_m, Y_m=L^{(1)}_m$, one can get exactly the algebra (\ref{3}). They form a closed subalgebra of $W_{1+\infty}$. The physical meaning of those generator is also clear: since the $L^{(n)}_m$ is a component of the conformal spin-(n+1) field, $T_m$ relates to spin-1 field and $Y_m$ spin-2 field. Notice that one can re-scale $L^{(0)}_m \rightarrow \alpha L^{(0)}_m$ without changing the algebra. It will be useful to relate $L^{(0)}_0$ to entropy later.

This $W_{1+\infty}$ has no central extension term, but one can also consider the central extension $W_{1+\infty}$ which has Kac-Moody algebra as subalgebra. The new algebra has the following commutation relation \cite{walg1},
\begin{equation}\label{5}\begin{split}
  [\tilde{L}^{(0)}_m, \tilde{L}^{(0)}_n]&=n\delta_{n+m,0},\\
  [\tilde{L}^{(1)}_m, \tilde{L}^{(0)}_n]&=-n \tilde{L}^{(0)}_{m+n},\\
  [\tilde{L}^{(1)}_m, \tilde{L}^{(1)}_n]&=(m-n)\tilde{L}^{(1)}_{m+n}+\frac{1}{12}n(n^2-1)\delta_{n+m,0},\\
    \cdots.
\end{split}\end{equation}
It contains an abelian Kac-Moody algebra and $c=1$ Virasoro algebra as a subalgebra.

Interesting this subalgebra also appears in Ref.\cite{ads1}, where a set of asymptotically $AdS_3$ boundary conditions were found. But the central charge and the level of the Kac-Moody algebra are different.
\section{Classification of black holes by the $W_{1+\infty}$ symmetry}
The $W_{1+\infty}$ symmetry was used to classify the universality classes of quantum Hall incompressible fluids \cite{ctz2}. These classes are specified by the kinematical data of the electrical charge $Q$ and the spin $J$ of the quasi-particles in quantum Hall fluid. They are the eigenvalues of $L^{(0)}_0$ and $L^{(1)}_0$, i.e.
\begin{equation}\label{6}
 L^{(0)}_0|Q>=e Q |Q>,\quad L^{(1)}_0 |Q>= J |Q>.
\end{equation}
All unitary, irreducible, highest-weight representation of the $W_{1+\infty}$ have been found \cite{kr1,kr2}. They exist for positive integer central charge $c=m=1,2,\cdots$ and are labeled by a $m-$component highest-weight vector $\vec{r}=\{r_1,\cdots,r_m\}$. For generic filling fractions of quantum Hall fluid, the spectrum of quasi-particles is given by
\begin{equation}\label{7}
  Q=\sum_{i,j=1}^m K_{ij}^{-1} n_j, \quad J=\frac{1}{2}\sum_{i,j=1}^m n_i K_{ij}^{-1} n_j,
\end{equation}
where $K_{ij}$ is an $m\times m$ symmetric, integral-valued matrix, and $n_j$ are integers represent the number of vortices (quasi-particles) created in the $j^{th}$ component of the fluid.

It was claimed that the black hole can be considered as a kind of topological insulator \cite{wangti1,wangti2}. So, in three dimensional spacetime, black holes can be considered as bilayer quantum Hall system, and we can classify the black holes with the help of $W_{1+\infty}$. For bilayer quantum Hall system, $c=m=2$, so are classified by two integer numbers $(n_1,n_2)$. The kinematical data $(Q,J)$ will be the eigenvalues of $L^{(0)}_0=T_0$ and $L^{(1)}_0=Y_0$, which relate to the entropy $S$ and angular momentum $J$ of the black holes \cite{bc1,wangwh1}. We denote the highest-weight state as $|Q,J>$ with
\begin{equation}\label{8}
 L^{(0)}_0|Q,J>=Q |Q,J>,\quad L^{(1)}_0 |Q,J>= J |Q,J>.
\end{equation}

Consider the BTZ black hole as an example. The eigenvalue of $T_0, Y_0$ were given \cite{nhs1}
\begin{equation}\label{9}
  T_0=\frac{\kappa r_+}{4 G},\quad Y_0=\frac{r_+ r_-}{4 G l},
\end{equation}
where $\kappa=\frac{r_+^2-r_-^2}{l^2 r_+}$ is the surface gravity and $l$ the AdS radius. To relate the $L^{(0)}_0$ with the entropy, we must re-scale this operator. To be consistent with the results of Ref.\cite{wangwh1}, we make the following choices,
\begin{equation}\label{9a}
 K=diag(2 k, -2 k), \quad Q=\frac{1}{\kappa k} T_0,
\end{equation}
where $k=\frac{l}{4G}$.

The relation (\ref{7}) give the following equations,
\begin{equation}\label{10}
  Q=\frac{1}{2 k}(n_1+n_2)=\frac{r_+}{l}=\frac{S}{2\pi k},\quad J=\frac{1}{4 k}(n^2_1-n^2_2)=\frac{r_+ r_-}{4 G l}.
\end{equation}

It is easy to get the solution
\begin{equation}\label{11}
  n_1=\frac{1}{4 G}(r_++r_-),\quad n_2=\frac{1}{4 G}(r_+-r_-).
\end{equation}
They are equivalent to the conditions
\begin{equation}\label{12}
  r_+=2(n_1+n_2)L_{PL}, \quad r_-=2(n_1-n_2)L_{PL},
\end{equation}
where $L_{PL}=G$ is the Planck length. It means that the radius of black hole is quantized.

The form of K-matrix (\ref{9a}) is similar to Laughlin states--the simplest fractional quantum Hall states. From this point of view, the BTZ black hole can be considered as the simplest fractional topological insulator.

In $W_{1+\infty}$ algebra, all operators $L^{(i)}_0$ are simultaneously diagonal and assign other quantum numbers to the quasi-particle \cite{ctz2}. In quantum Hall fluid, those quantum numbers measure the radial moments of the charge distribution of a quasi-particle. In black hole physics, those quantum numbers can be considered as `W-hairs' and maybe crucial for information paradox \cite{whair1}.
\section{Conclusion}
In this paper, we consider the relation of the near horizon symmetry algebra of black holes in three dimensional spacetime with the $W_{1+\infty}$ symmetry algebra of quantum Hall effect. It is found that the former is a subalgebra of the latter. The $W_{1+\infty}$ algebra is a quantum version of area-preserving diffeomorphism algebra, which is a \emph{dynamical symmetry} of quantum Hall liquid. This is because the quantum Hall droplet is incompressible fluid, so has constant area. If we consider black hole as quantum hall droplet, the origin of the near horizon symmetry is very clear: it comes from the area-preserving diffeomorphism. With the help of the $W_{1+\infty}$ algebra we can also give classification of black holes. They are specified by two integer $(n_1,n_2)$ which are functions of the black hole entropy and angular momentum $(S,J)$. 

Now let's pay attention to four dimensional spacetime. Due to the above discussion, the near horizon symmetry algebra of black hole can be considered as sub-algebra of the volume-preserving differomorphism. The representation of this algebra may also give some classification of black holes in four dimensional spacetime. Interestingly, the volume-preserving differomorphism algebra and the related D-algebra appears at the higher dimensional topological insulators \cite{dalgebra}.

The central extension of the supertranslation algebra is an abelian Kac-Moody algebra. This algebra can also be get from another approach. It is well known that, $(2+1)-$dimensional general relativity with $\Lambda=-\frac{1}{l^2}$ can be cast into $SO(2,1)\times SO(2,1)$ Chern-Simons theory. On a manifold with a boundary, the Chern-Simons theory reduces to a chiral Wess-Zumino-Novikov-Witten(WZNW) theory with $SO(2,1)$ Kac-Moody algebra on the boundary. If the boundary is chosen to be the horizon of a black hole, the $SO(2,1)$ Kac-Moody algebra reduces to $SO(1,1)$ Kac-Moody algebra \cite{whcft1}.

$W_{1+\infty}$ is also used to explain the Hawking radiation \cite{hawkradia1,hawkradia21,hawkradia2,hawkradia3,hawkradia4,hawkradia5} and solve the information paradox \cite{whair2,whair3,whair4,whair5}, since it has infinite set of quantum numbers, i.e. `W-hair'. Due to our result, the near horizon symmetry algebra is a subalgebra of $W_{1+\infty}$, so the `W-hair' can be considered as generalized `soft hair' \cite{shairs}. Since infinite-dimensional algebra appears rarely in physical system, the relation between those two algebra should not be a coincide. It gives another evidence that our early claim ``black hole can be considered as a kind of topological insulator" is correct. This claim relate the black hole physics with the condensed matter physics. It is also the starting point to relate the gravity with some non-trivial condensed matter systems \cite{hu1,volo1,lib1,vaid1,highd1}.

\acknowledgments
 This work is supported by the NSFC (Grant No.11647064) and Nanhu Scholars Program for Young Scholars of XYNU.

\bibliography{bms1}
\end{document}